\documentclass{webofc}
\usepackage[varg]{txfonts}   
\usepackage{listings}
\usepackage{hyperref}
\begin{document}
\title{Spitzer Observations of Large Amplitude Variables in the LMC and IC\,1613}
%
%

\author{\firstname{Patricia A.} \lastname{Whitelock}\inst{1}
\fnsep\thanks{\href{mailto:paw@saao.ac.za}{\tt paw@saao.ac.za}  } 
\and
        \firstname{Mansi} \lastname{Kasliwal}\inst{2}
 \and
        \firstname{Martha} \lastname{Boyer}\inst{3,4}     
}

\institute{South African Astronomical Observatory \& Astronomy Department,  University of Cape Town, South Africa  
\and
           Division of Physics, Mathematics and Astronomy, CIT, Pasadena, CA91125, USA
\and
NASA Goddard Space Flight Center, MC 665, 8800 Greenbelt Road, Greenbelt, MD 20771 USA
\and
Department of Astronomy, University of Maryland, College Park, MD 20742 USA
          }
\abstract{The 3.6 and 4.5 $\mu$m characteristics of AGB variables in the LMC and IC\,1613 are discussed.  For C-rich Mira variables there is a very clear period-luminosity-colour relation,  where the $[3.6]-[4.5]$ colour is associated with the amount of circumstellar material and correlated with the pulsation amplitude. The [4.5] period-luminosity relation for dusty stars is approximately one mag brighter than for their naked counterparts with comparable periods.
}
\maketitle

\section{Introduction}\label{sec:intro}
The material discussed here involves observations made with the Spitzer spacecraft at 3.6 and 4.5 $\mu$m, and comes from the SAGE Var Collaboration \cite{riebel} for the LMC and from the SPIRITS \cite{kasliwal} and DUSTiNGS  \cite{boyer} collaborations for IC\,1613. We focus on the large amplitude AGB stars, which were found by SPIRITS as a byproduct of their search for infrared (IR) transients and by DUSTiNGS because of their strong dust emission. 

These stars have pulsation periods that are mostly in the range 100 to  1000 days, and peak-to-peak amplitudes over 0.3 mag at [3.6] and/or [4.5]. They are very luminous bolometrically and in the IR. They are near the very top of the AGB,  so they represent the brightest evolutionary phase for stars of a particular mass/metallicity. They are bright in the IR because they are cool and because they become enshrouded in dust during their AGB evolution, so they are particularly bright at the wavelengths of interest to JWST and the next generation of extremely large ground-based telescopes. The variables with the largest amplitudes are traditionally known as Miras, although those with very thick dust shells are often identified as "extreme-AGB" (x-AGB) stars (often, but not necessarily, C-rich),  or as OH/IR stars (if they are O-rich). 

The Miras are potentially useful distance indicators (rivalling Cepheids) \cite{feast3} as well as for studies of resolved old and intermediate mass populations in a range of galaxies. 
They are astrophysically important because they are responsible for some of the processed material returned to the interstellar medium, including important elements such as carbon, lithium and elements produced by the s-process. The details of this are still rather sketchy because the process of mass loss is not yet well understood. 

AGB variables in the Magellanic Clouds fall into three classes \cite{soszynski}: Miras, semi-regular variables (SRVs) and OGLE small amplitude red variables (OSARGS), of which only the Miras and SRVs have sufficiently large amplitudes to be of interest here. Wood et al.\ \cite{wood} noted that AGB stars occupied various period-luminosity (PL) relations. Miras pulsate in the fundamental mode{\footnote{possibly hot bottom burning Miras pulsate in the first overtone \cite{feast4}.} while SRVs pulsate in the fundamental and/or in overtones. The Miras generally show only a single mode while SRVs often show two or more modes.  Most Miras have larger amplitudes than most SRVs, but the two types are often difficult to distinguish without a large number of observations \cite{soszynski}. 

\begin{figure}
\centering
\includegraphics[width=6cm,clip]{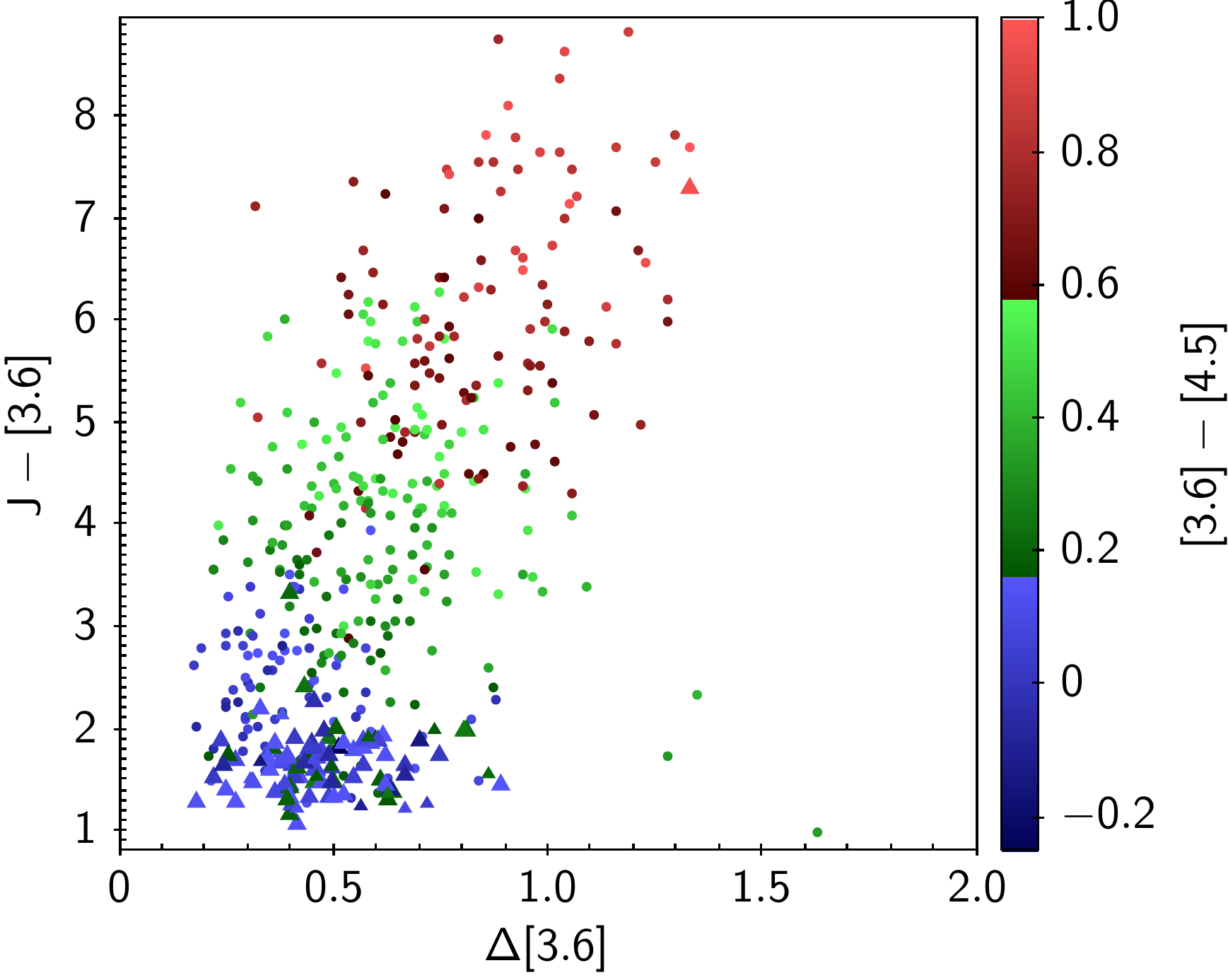}
\caption{Colour ($J-[3.6]$) against amplitude at $3.6\ \mu$m ($\Delta [3.6]$). C- and O-rich stars (as determined from SED fits \cite{riebel}) are shown as circles and triangles, respectively. Note that the redder C-stars tend to have larger amplitudes.}
\label{fig1}       
\end{figure}
\section{Large Magellanic Cloud (LMC)}\label{LMC}
Riebel et al.\ \cite{riebel} obtained observations of the LMC bar at 3.6 and $4.5\mu$m at four epochs, which they combined with the two earlier epochs of the SAGE legacy survey to give a total of six epochs. There are  good periods for these variables from the OGLE III survey \cite{soszynski} at $V$ and/or $I$, and single epoch $JHK_S$ photometry from the InfraRed Survey Facility (IRSF) in South Africa \cite{kato}. 

Figure \ref{fig1} shows a plot of $J-[3.6]$ against $\Delta[3.6]$. For the O-rich stars, there is no clear correlation of amplitude with colour, while the C-stars show a relation with redder stars having larger amplitudes.  Of course with just six epochs  $\Delta[3.6]$ is only a lower limit and the $J$ measurement comes from a single epoch, hence the large scatter.  The difference between the O- and C-rich stars may be related to the relative transparency of the O-rich grains \cite{bladh}. Riebel et al.\ analysed the Spitzer data, together with periods from MACHO, to establish PL relations for Mira variables with $J-[3.6]<3.1$ only, i.e. excluding the redder  x-AGB stars. 

\begin{figure*}
\centering
\includegraphics[width=10cm,clip]{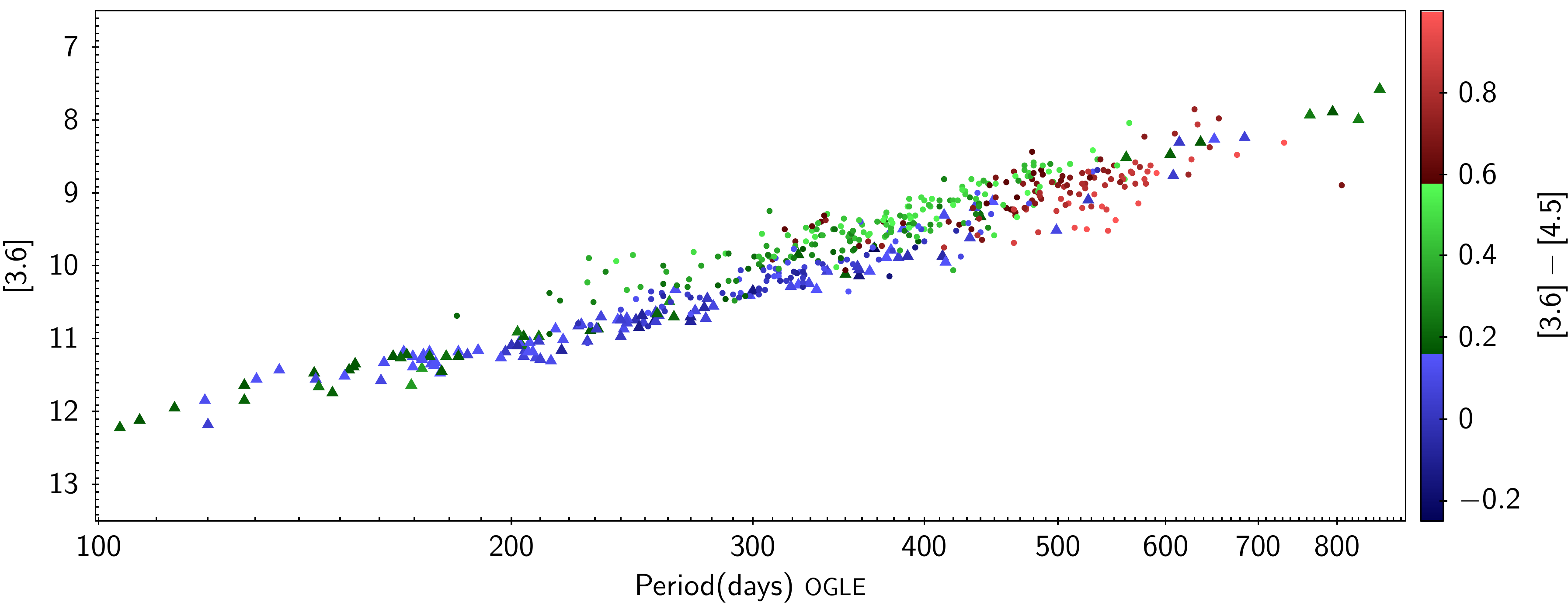}
\includegraphics[width=10cm,clip]{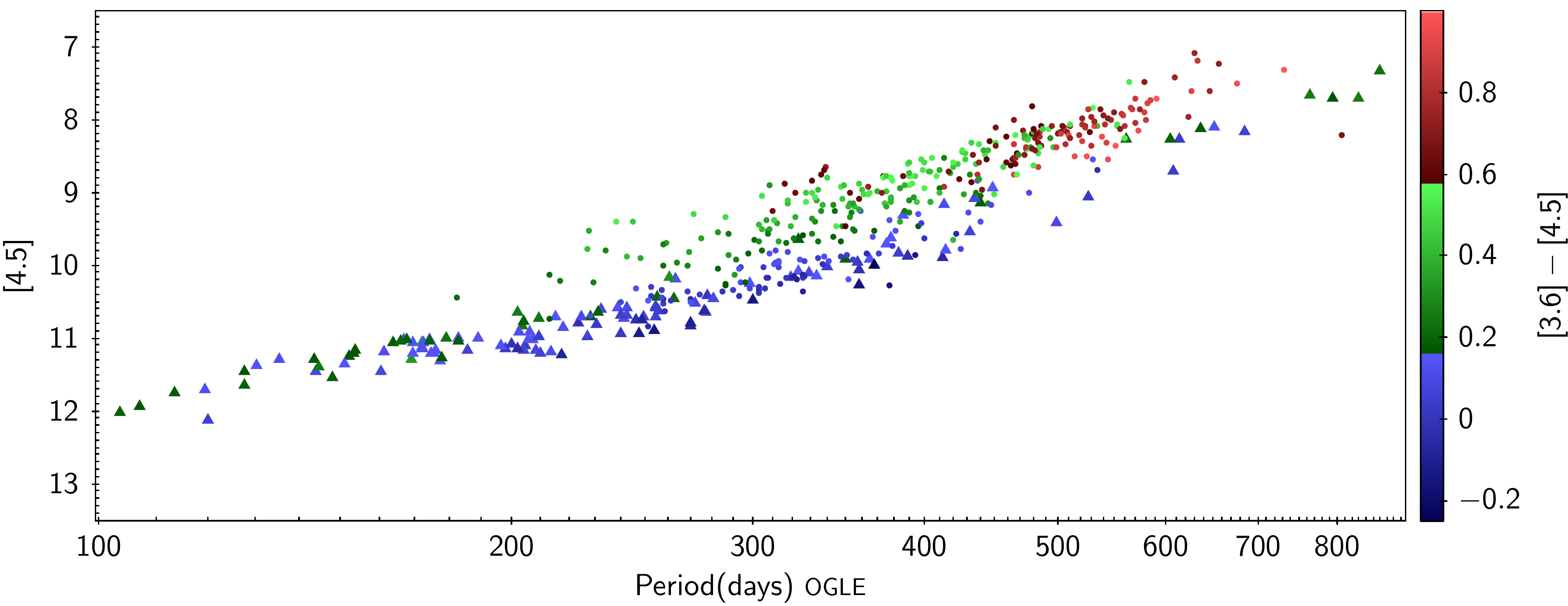}
\caption{The [3.6] and [4.5] PL relations for Spitzer variables in the LMC that are classified as Miras by OGLE. O-rich and C-rich stars are identified as triangles and circles respectively, while the colour coding is according to $[3.6]-[4.5]$ as shown on the right-hand side. }
\label{fig2}       
\end{figure*}

Figure \ref{fig2} shows the [3.6] and [4.5] PL relations for all of the variables that the OGLE group \cite{soszynski} identify as Miras. OGLE do not distinguish variability class on the basis of colour and this therefore includes the redder x-AGB stars as well as those Riebel et al.\ \cite{riebel} considered.  
Riebel et al.\ identified only 10 AGB candidates in their Spitzer selected variables, that were not previously known variables. These are probably Miras or SRVs that were either not observed or not classified by OGLE, possibly because they were too red.  We can therefore be confident that the sample of Miras considered here is very nearly complete. 

 Riebel et al.\ discussed the PL relations for O- and C- rich stars and noted the surprisingly large scatter, particularly in the [4.5] relation; by including the x-AGB stars we see even more spread.
Some of this may be due to variations in the CO fundamental vibration-rotation feature which falls within the [4.5] band, but it is clear from the colour dependence that most of it is due to dust emission. The spread at given period is over one magnitude and the brighter sources have larger $[3.6]-[4.5]$ indices.  This is very clear for the C-rich Miras in Figure \ref{fig3}, where the [4.5] relation is contrasted with the $K_S$ relation. At [4.5] the reddest stars are the brightest because of dust emission, while at $K_S$ they are the faintest because of dust absorption. It is also notable in this figure that the spread in $K_S$ of the reddest stars is around three mag whereas that in [4.5] is less than one mag. This is partially because the $K_S$ observations are single epoch,  but the asymmetric character of the dust shells around C-stars must also contribute to this scatter. There is a good deal of evidence that C-rich Miras eject dust in random directions 
  (\cite{feast}, \cite{whitelock2}), in the same way as RCB stars, possibly because the dust forms over convection cells \cite{feast2}.  The dust will be optically thin at [3.6] and [4.5] so the entire shell will be measured in those bands.

\begin{figure*}
\centering
\includegraphics[width=0.49\linewidth]{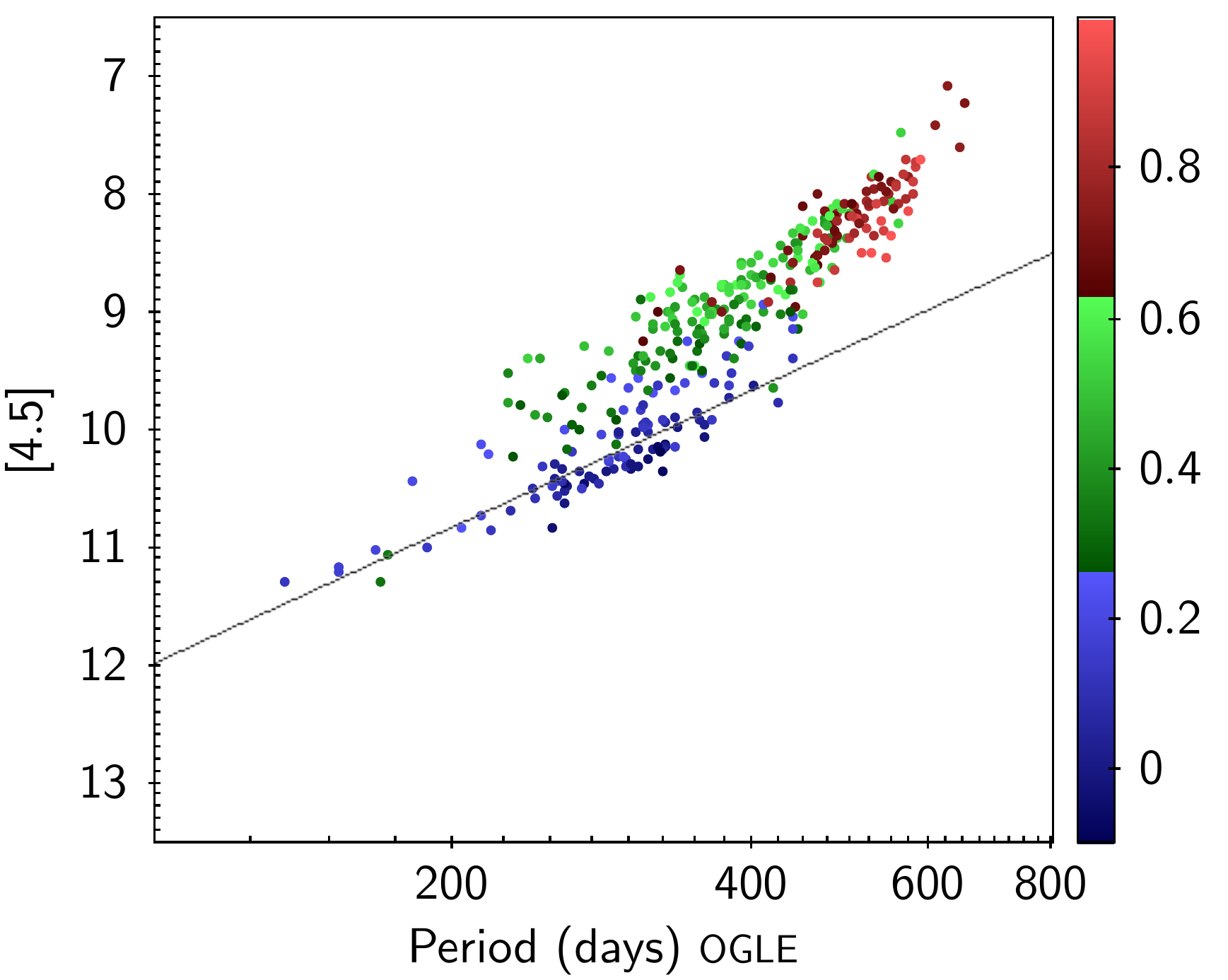}
\hfill
\includegraphics[width=0.49\linewidth]{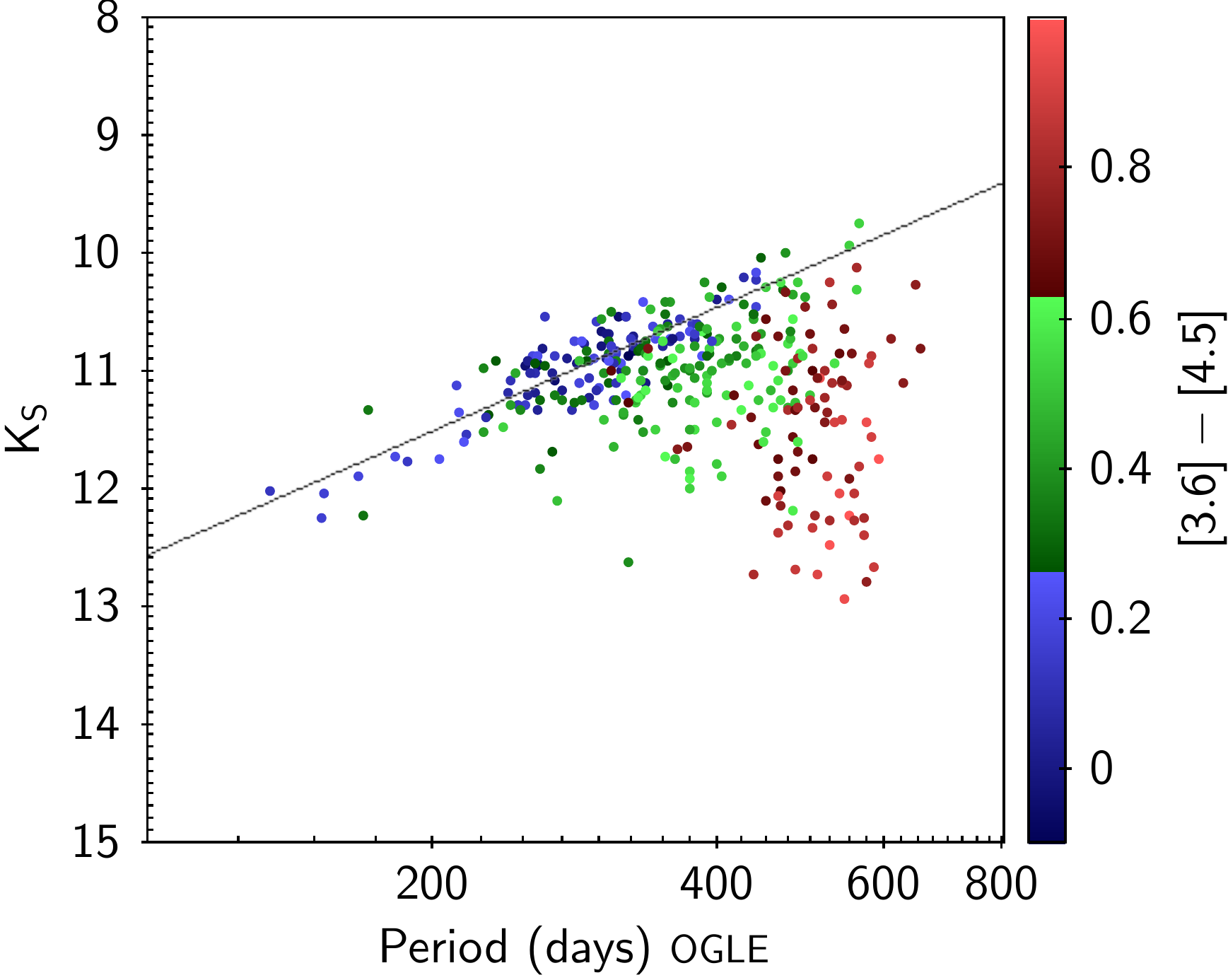}
\caption{PL relationships for the same C-rich stars at [4.5] and at $K_S$. The straight lines are the Mira PL relationships from Riebel et al.\ \cite{riebel} (derived from the bluer group of this sample of stars)  and Whitelock et al.\ \cite{whitelock} (for a completely different sample of stars).}
\label{fig3}       
\end{figure*}

Riebel et al.\ colour-selected  Miras for their analysis so that most of their 3 to 4 $\mu$m flux originated from the star rather than from the shell.  The spread we see in the [4.5] PL relation (Figure \ref{fig3} left panel)  is due to the increasing contribution from dust as the star gets redder, i.e. there is a period-luminosity-colour relation for these stars. It is particularly interesting to see that once the dust shell dominates the flux the period-luminosity relation reestablishes itself, almost parallel to the original relation, but about 1.2 mag brighter than that seen for the naked stars. The effect is much smaller at [3.6], because the contribution from dust emission is weaker. It will be interesting to examine the bolometric luminosities of these variables and to compare with models to see if those with thicker shells can be understood as more evolved than those without.  There is evidence that Miras do not change their periods dramatically once they start pulsating in the fundamental mode (e.g.\cite{feast3}), but some change in period may occur as the star evolves through the instability strip \cite{wood1}.  Only stars with a limited range of initial masses will become C-rich as hot bottom burning (HBB) will keep stars over   3 or 4 $M_{\odot}$ O-rich.


\section{IC1613}\label{ic1613} 
Menzies et al.\ \cite{menzies} discussed the variable AGB stars in IC\,1613, identifying several O- and C-rich Miras with periods ranging from 260 to 880 days from $JHK$ photometry. Of particular interest were four O-rich Miras, with periods ranging from 460 to 580 days and luminosities above the $K_S$ and $m_{bol}$ PL relations. The detection of lithium in one of these demonstrated that it is undergoing HBB, and we can presume that this is why these stars depart from the PL relation (which is linked to the core-mass luminosity relation at the tip of the AGB). This survey was sensitivity limited and would not have detected the most dusty variables, which are faint at $J$.

The SPIRITS and DUSTiNGS surveys do not cover exactly the same region of IC\,1613, but they recover those variables found by \cite{menzies} in the overlap region, and detect several more that are particularly dusty. Although we can determine periods for many of these, there are insufficient observations to distinguish SRVs from Miras. Therefore, in Figure \ref{fig4}, we compare them  with OGLE and Spitzer data from the LMC for both SR and Mira variables. We  distinguish large  ($\Delta [3.6]$ and $\Delta [4.5] > 0.5$ mag) and small pulsation amplitudes, noting that those with large amplitudes are almost certainly Miras.  Within the LMC most of the large amplitude variables have periods over 300 days and are more luminous than their low amplitude counterparts at a given period, although there are a few large amplitude stars at shorter periods and lower luminosities. The variables in IC\,1613, which have periods from 180 to 580 days, follow very much the same trends. There are a few of the lower amplitude stars that appear to be overtone pulsators. The four HBB stars cluster closely together as they do at shorter wavelengths; they are amongst the brightest variables at $3.6 \mu$m, but are are not as red as the most luminous C-stars, which have comparable periods.

\begin{acknowledgement} 
\noindent\vskip 0.2cm
\noindent {\em Acknowledgments}: This work makes use of VizieR, the CDS catalogue service,  and the TOPCAT\cite{taylor} data processing package. PAW acknowledges the receipt of a research grant from the South African National Research Foundation (NRF).
\end{acknowledgement}

\begin{figure*}[t]
\centering
\includegraphics[width=0.49\linewidth]{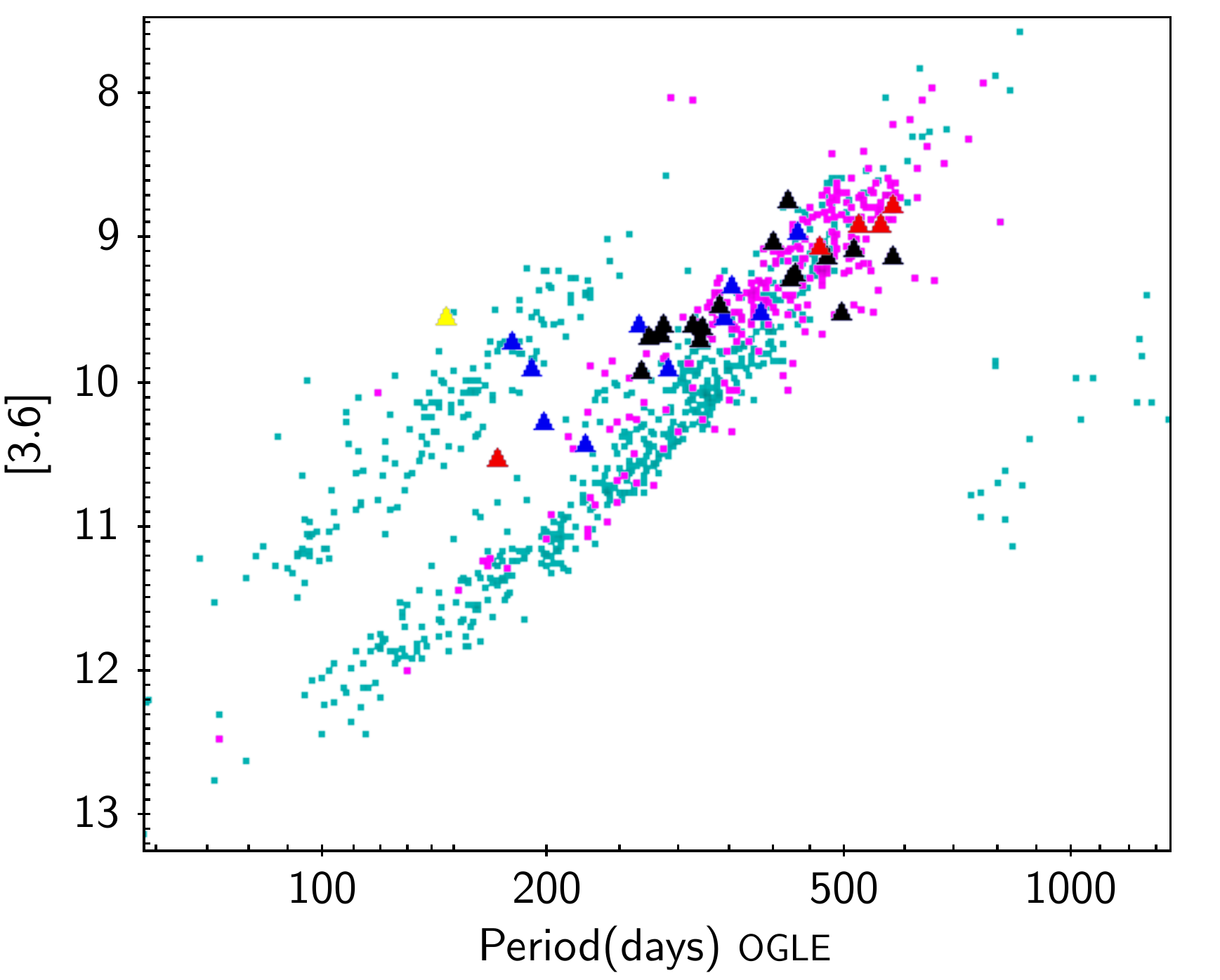}
\hfill
\includegraphics[width=0.49\linewidth]{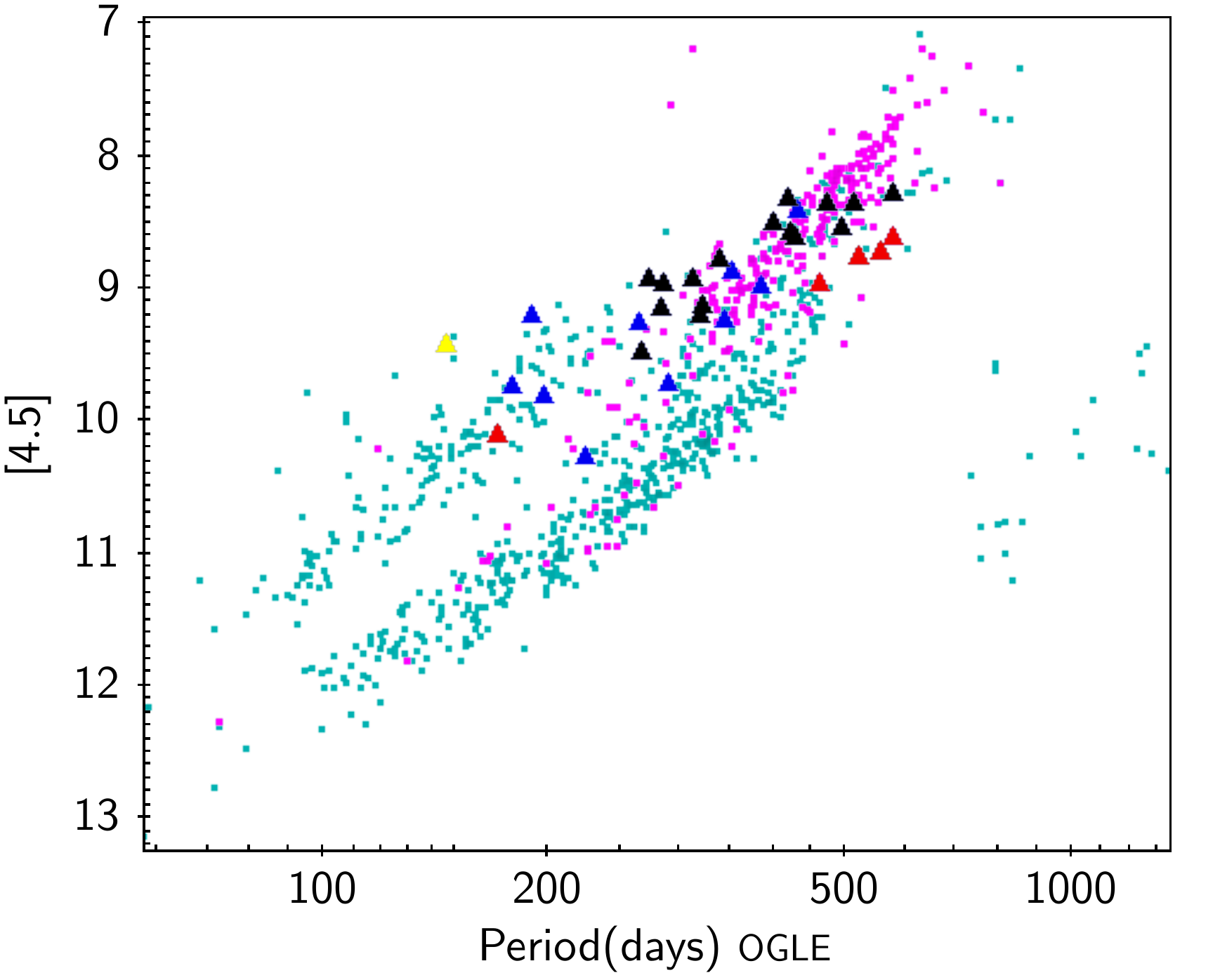}
\caption{[3.6] and [4.5] PL relations for the SR and Mira variables in the  LMC (small circles: cyan, except for stars with $\Delta [3.6]$ and $\Delta [4.5] > 0.5$ mag, which are magenta), and in IC\,1613 (triangles: O-rich stars in red and presumed C-rich stars in blue or if  $\Delta [3.6]$ and $\Delta [4.5] > 0.5$ mag, in black; the single yellow symbol is a Cepheid).  The magnitudes of the IC\,1613 variables have been corrected by 5.83 mag, for the relative distances of the two galaxies. The two cyan sequences that stand out show the fundamental and first overtone pulsators. }
\label{fig4}       
\end{figure*}

%
%

\end{document}